\documentclass[11pt,a4paper]{article}

\usepackage{a4wide}
\usepackage{graphicx}
\usepackage[english]{babel}
\usepackage{amsmath}
\usepackage{amssymb}
\usepackage{epsfig}

\newcommand{\myincludegraphics}[2][1]{\includegraphics[#1]{#2}}

\newcommand{\qed}{\hfill \ensuremath{\Box}}

\newtheorem{theorem}{Theorem}
\newtheorem{lemma}{Lemma}

\newenvironment{proof}[1][{\em Proof.}]{\begin{trivlist}
  \item[\hskip \labelsep {\bfseries #1}]}{\end{trivlist}}

\newcommand{\ie}{{\em i.e.}}

\newcommand{\cns}
{complex networks}

\newcommand{\cn}
{complex network}

\newcommand{\network}
{graph}

\newcommand{\networks}
{graphs}

\newcommand{\link}
{edge}

\newcommand{\links}
{edges}

\newcommand{\nodes}
{vertices}

\newcommand{\node}
{vertex}

\newcommand{\cc}
{clustering}

\newcommand{\Actors}{{\em Actors}}
\newcommand{\Internet}{{\em Internet}}
\newcommand{\Web}{{\em Web}}
\newcommand{\Cooccurrence}{{\em Cooccurrence}}
\newcommand{\Coauthoring}{{\em Coauthoring}}
\newcommand{\Proteins}{{\em Proteins}}

\begin{document}


\title{Bipartite Graphs as Models of Complex Networks}
\author{Jean-Loup Guillaume and Matthieu Latapy \\
        \textsc{liafa} -- {\sc cnrs} -- Universit\'e Paris 7\\
        2 place Jussieu, 75005 Paris, France.\\
        (guillaume,latapy)@liafa.jussieu.fr}
\date{}
\maketitle

\vskip 0.1cm

\noindent

\begin{abstract}
It appeared recently that the classical random graph model used to
represent real-world \cns\ does not capture their main
properties. Since then, various attempts have been made to provide
accurate models. We study here a model which achieves the
following challenges: it produces graphs which have the three main
wanted properties (clustering, degree distribution, average distance),
it is based on some real-world observations, and it is sufficiently
simple to make it possible to prove its main properties.  This model
consists in sampling a random bipartite graph with prescribed degree
distribution. Indeed, we show that any \cn\ may be viewed as a
bipartite graph with some specific characteristics, and that its main
properties may be viewed as consequences of this underlying
structure. We also propose a growing model based on this observation.
\end{abstract}


\vspace*{-0.3cm}

\section*{\label{sec_intro}Introduction.}

It has been shown recently that most real-world \cns\ have some
specific properties in common. These properties are not captured by
the model generally used before this discovery, although they play a
central role in many contexts like the robustness of the Internet
\cite{albert00error,Callaway00network,cohen00resilience,cohen01breakdown,motter02cascade},
the spread of viruses or rumors over the Internet, the Web or other
social networks \cite{MN00,New02,PV01}, the performance of
protocols and algorithms
\cite{kim02path,magoni01influence,walsh99search}, and many other.

This is why a strong effort has been put in the realistic modeling of
\cns\ in the last few years,
and much progress has been accomplished in this 
field. Some models achieve the aim of producing \networks\ which
capture some, but not all of the main properties of real-world
\cns. Some models obtain all the wanted properties but rely on
artificial methods which give unrealistic graphs (trees, graphs with
uniform degrees, etc).  Others rely on construction processes which
may induce some hidden properties, or are too difficult to analyze.

\medskip

In this paper, we propose the random bipartite graph model as a
general model for \cns. We will show that this model produces graphs
with many observed properties. It relies on real-world observations
and gives realistic graphs. Finally, it is simple enough to make it
possible to prove its main properties. We will also discuss some
identified drawbacks of this model.

We will first present an overview of the context in which our work
lies. In particular, we use some ideas introduced in previous papers,
which we need to describe precisely. Then we show how {\em all} \cns\
may be described as bipartite structures. After this, we present the
random bipartite model and analyze it to show that the main properties
of \cns\ are somehow a consequence of their underlying bipartite
structure. We also present a growing bipartite model based on the same
ideas. Finally we discuss the advantages and limitations of these models.

\section{\label{sec_context}Context.}

Throughout our presentation, we will use a representative set of \cns\
which have received much attention and span quite well the variety of
contexts in which \cns\ appear: an Internet graph at router
level~\cite{chen02origin,mercatorurl,govindan00heuristics} consisting
of physical links between routers ; a web graph from Notre Dame
university~\cite{albert99diameter,notredameurl} where web pages are
linked by hyperlinks ; a co-occurence graph in which words are linked
if they belong to a same sentence in a given text
\cite{ferrer01small,bibleurl} ; the actor graph where actors are linked
when they have played together~\cite{imdburl,watts98collective} ; a
co-authoring graph from Arxiv~\cite{arxivurl} where scientists are 
linked if they have signed a paper
together~\cite{newman01scientific1,newman01scientific2,newman01random};
and finally a protein graph where two proteins of a given biological
system are linked if they influence  each
other~\cite{notredameurl,jeong00largescale}.

Many other \cns\ have been studied.
Refer to
\cite{albert02statistical,dorogovtsev02evolution,newman03structure}
for a more descriptive list of networks and corresponding references.
All these networks have some properties in common which have been
discovered quite recently and have concentrated a large attention in
various communities. Hereafter we present the properties in concern
and some recent efforts in the modeling of these properties.


\subsection{Statistical properties}

Most real-world \cns\ have a number of \links\ $m$ which scales
linearly with the number of \nodes\ $n$: $m \sim k\cdot n$ where $k$
is the average degree (which does not depend on the size of the
\network). Therefore, these networks have a low density (going to $0$
when $n$ grows), the density being defined as the number of
existing \links\ over the number of \links\ that could exist
($n(n-1)/2$).


The distance between two \nodes, defined as the number of \links\ on a
shortest path between these \nodes, is low on average. It is a well
known property on social networks since the work of Stanley Milgram
\cite{milgram67small} and the notion of ``six degrees of
separation''. However it appeared more recently that all \cns\ have a
low average distance which typically scales like
the logarithm of the size of the graph. It has been shown that this is
actually true for any graph which contains some reasonable amount of
randomness. Actually, under reasonable  assumptions, the average
distance in random graphs  scales  even slower than the
logarithm\,\footnote{Typically like $\log(n)/\log(\log(n))$.} of their
size
\cite{bollobas85random,cohen02structural,cohen03ultrasmall,dorogovtsev03metric,lu01diameter,newman01arbitrary,newman01random}.

The local \cc, or clustering coefficient \cite{watts98collective} is
defined for each \node\ of degree at least $2$ as the
proportion of \links\ between its neighbors: $c(u)=\frac{|\{(x,y), x,y\in
N(u)\}|}{{d(u) \choose 2}}$, where $d(u)$ is the degree of \node\ $u$
and $N(u)$ is the set of neighbors of $u$. The \cc\ of a graph is
simply the average over all \nodes.
All real-world \cns\ have a high \cc\ which
seems to be independent of the size of the network. These \networks\
are locally dense while globally sparse.

Finally, the degree distribution which is, for each $k$, the
probability $p_k$ that a randomly chosen \node\ has degree $k$, is
completely different from what was expected. Indeed for almost all
real-world \cns, the degree distribution follows a power law: $p_k\sim
k^{-\alpha}$, while one would have expected an exponential decrease
(Poisson-like distributions). The exponent $\alpha$ of the power law
is generally between $2$ and $3$. Such a distribution means that
although most \nodes\ have a small degree, the number of \nodes\ with
degree $k$ decays only polynomially with $k$, and therefore there is a
significant number of \nodes\ with high degree.

\medskip

The main properties of the real-world \cns\ we use in this paper are
given in Table~\ref{tab_gr_reels}. Notice that, as announced, all
these real-world \cns\ have a very low density, a low average
distance, a power law distribution of degrees and a high clustering.

\begin{table}
\begin{center}
\small
\begin{tabular}{|l|c|c|c|c|c|c|c|} \hline
       & Internet  & Web     & Actors   & Co-auth  & Co-occur & Protein  \\ \hline \hline
$n$      & 75885   & 325729  & 392340   & 16401    & 9297     & 2113     \\ \hline
$m$      & 357317  & 1090108 & 15038083 & 29552    & 392066   & 2203     \\ \hline
$density$& 1.2e-4  & 2.1e-5  & 1.9e-4   & 2.2e-4   & 9.1e-3   & 9.9e-4\\ \hline
$c$      & 0.171   & 0.466   & 0.785    & 0.638    & 0.822    & 0.153    \\ \hline
$\alpha$ & 2.5     & 2.3     & 2.2      & 2.4      & 1.8      & 2.4      \\ \hline
$d$      & 5.80    & 7       & 3.6      & 7.18     & 2.13     & 6.74     \\ \hline
\end{tabular}
\caption{\label{tab_gr_reels} The main statistics for the complex
networks we use in this paper. For each network, we give its number of
vertices $n$, its number of links $m$, its density, the value of the
exponent $\alpha$ of the power law that fits best its
degree distribution, its clustering
$c$, and its average distance $d$.
}
\end{center}
\end{table}

The similarity of these networks concerning unexpected properties has
led to the study of other properties. The simplest one concerns the
degree-degree correlation: what is the average degree of the neighbors
of a \node\ of degree $k$. Three main behaviors are expected, either
high-degree \nodes\ tend to connect to high-degree \nodes, or to
low-degree \nodes, or to any nodes. These behaviors can be observed
using the the slope (increasing, decreasing or constant) of the plot
which relates the average degree of the neighbors of nodes of degree
$k$, to $k$
\cite{boguna03epidemic,subramanian02characterizing,tangmunarunkit01network},
or with a single parameter (assortativity coefficient), which may be
positive (assortative networks), negative (dissortative networks) or
null (neutral networks)~\cite{newman03mixing}. Most social networks
are assortative (similar \nodes\ are connected) while technological or
biological are generally dissortative~\cite{newman03mixing}.

One may also correlate the clustering and the degree by computing the
average \cc\ of \nodes\ with a given degree. This also defines
assortative (high degree yields high clustering), neutral or
dissortative networks.

Finally, other properties have been studied, such as the centrality
(how many shortest paths contain a given \node)
\cite{newman01scientific2}, the distribution of eigenvalues of the
adjacency matrix \cite{faloutsos99powerlaw,mihail02eigenvalue}, 
etc. All these statistical properties are used to describe a given
\cn\ and to study the similarities and differences between several
\cns. They give precise insight on what one may expect when considering
a \cn\ having a set of properties. 


\subsection{Modeling \cns}

The basic model for \cns\ is the Erd\"os-R\'enyi (ER) random graph
model \cite{bollobas85random,erdos59random}. In a random graph with
$n$ \nodes, each of the $\frac{n\cdot (n-1)}{2}$ possible \links\
exists with a given probability $p$ (this model is know as ${\cal
G}_{n,p}$).
The average distance of ${\cal G}_{n,p}$ scales with the
logarithm of $n$ \cite{bollobas85random}. Moreover, the \cc\ is equal
to the connection probability $p$ since each pair of \nodes\ is
connected with the same probability independently of the fact that
they are both linked to a same \node. If $m \sim k\cdot n$ as in
real-world \cns, this means that the \cc\ scales as $n^{-1}$ and
therefore tends to $0$ when $n$ grows. Finally, the degree
distribution follows a Poisson law $p_k \sim
e^{-\lambda}\frac{\lambda^k}{k!}$ \cite{bollobas85random}, which implies
in particular that all \nodes\ have nearly the same degree.

Therefore, although this model can be considered as relevant
concerning the average distance, it misses two main properties of
real-world \cns. In particular, the degree distributions 
are qualitatively different.

\medskip

It is however possible to sample uniformly a random graph with a given
degree distribution (in particular a power law)
\cite{bender78asymptotic,bollobas80,luczak92sparse,molloy95critical,molloy98size}
using the configuration model, or Molloy
and Reed\,\footnote{Despite it has been introduced
in~\cite{bender78asymptotic} and studied in~\cite{bollobas80}, this
model is often referred to as the {\em Molloy and Reed} model
since these authors made it popular in their contributions
\cite{molloy95critical,molloy98size}. We will follow this convention
here.} (MR) model: for each \node, draw its degree at random
according to the given distribution, create as many connection points
as its degree and finally connect pairs of connection points at
random\footnote{Note that this algorithm may induce multiple links
and loops.
One may
also use techniques to avoid them, see for instance~\cite{latapy05random,milo03uniform},
but this is out of the scope of this paper.}.

The power law graphs obtained this way have an average
distance which scales slower that the logarithm of their size
\cite{cohen02structural,cohen03ultrasmall,dorogovtsev03metric,lu01diameter,newman01arbitrary,newman01random}.
Moreover, the fact that \nodes\ are
linked together purely at random (only their number of \links\ is
given) makes it possible to study the properties of the obtained
graphs, and it indeed seems that it captures some of the most important
behaviors of \cns\
\cite{aiello00random,cohen02structural,lu01diameter,newman03structure,PV01}.
However, under reasonable assumptions on the degree distribution, the
\cc\ of these graphs tends to zero when $n$ grows \cite{newman03structure}.


\medskip

This approach could in principle be
continued to sample a graph among the ones having a given number of
nodes, a given degree distribution {\em and} a given \cc. However,
until now, there is no known method to sample uniformly such a graph,
and the problem seems difficult.

\bigskip

On the other hand, a large variety of models based on the iteration of
a construction process {\em inspired} from the way \cns\ grow in
reality have been introduced.

The first generic model of real-world \cns, has been introduced in
1998 by Watts and Strogatz (WS) \cite{watts98collective}. One starts
with a ring of $n$ \nodes\ in which each \node\ is connected to its
$k$ nearest neighbors, for a given $k$. Then, each \link\ is rewired
with a given probability $p$ by choosing randomly a new
extremity. When $p$ is small the \network\ is almost a ring which have
high average distance and high clustering. On the opposite, when $p$
is high, the \network\ is nearly random. For medium values of $p$, the
\network\ has both a small average distance and a high \cc, but the
degree distribution of the obtained \networks\ does not follow a power
law~\cite{dorogovtsev00exactly,watts98collective}.



Another important step was done with models based on {\em preferential
attachment}.
\cite{barabasi99emergence,dorogovtsev00structure,kumar-stochastic}.
The ``rich gets richer'' principle can be derived in a model where
\nodes\ arrive one by one in a \network\ and choose their neighbors
with a probability proportional to the degree of these neighbors. This
model has been studied intensively and is now well
known~\cite{albert02statistical}: the degree distribution follows a
power law with exponent, the average distance is logarithmic in the
number of \nodes, and the \cc\ is low, going to $0$ when the number of
\nodes\ grows. Despite this last point, this model has received much
attention, in particular because it defines {\em growing} graphs.

Both the WS model and the AB one have been introduced to model generic
behavior of \cns. If they both fail in producing \networks\
having each of the three properties we cited, 
they have been widely used as building blocks for more complex
models.

\medskip

Table~\ref{tab_models} shows the performances obtained with the basic
models we cited in our practical cases. Let us insist on the fact that
the models seek {\em qualitative} properties (non negligible \cc,
power law degree distribution, etc). Their aim is not to produce
graphs with exactly given values for these properties. However, even
with this in mind, the graphs obtained using these models are
significantly different from real-world ones concerning at least one
of these three points.

\begin{table}[!h]
\begin{center}
\small
\begin{tabular}{|l|c|c|c|c|c|c|c|} \hline
       & Internet  & Web     & Actors   & Co-auth  & Co-occur & Protein  \\ \hline \hline
$n$      & 75885   & 325729  & 392340   & 16401    & 9297     & 2113     \\ \hline
$m$      & 357317  & 1090108 & 15038083 & 29552    & 392066   & 2203     \\ \hline

$c$      & 0.171   & 0.466   & 0.785    & 0.638    & 0.822    & 0.153    \\ \hline
$c_{ER}$ & 0.0001  & 0.00002 & 0.0002   & 0.0002   & 0.009    & 0.001    \\ \hline
$c_{MR}$ & 0.0694  & 0.017   & 0.0057   & 0.001    & 0.26     & 0.007    \\ \hline
$c_{AB}$ & 0.0024  & 0.0005  & 0.0015   & 0.003    & 0.028    & 0        \\ \hline
$c_{WS}$ & 0.171   & 0.461   & 0.74 (*) & 0.523 (*)& 0.74 (*) & 0.06 (*) \\ \hline \hline

$d$      & 5.80    & 7       & 3.6      & 7.18     & 2.13     & 6.74     \\ \hline
$d_{ER}$ & 5.25    & 5.47    & 2.97     & 7.57     & 2.06     & 10.4     \\ \hline
$d_{MR}$ & 3.25    & 4.48    & 2.95     & 5.77     & 2.36     & 5.73     \\ \hline
$d_{AB}$ & 4.15    & 5.1     & 2.93     & 5.5      & 2.38     & 8.15     \\ \hline
$d_{WS}$ & 5.90    & 11.23   & 2559 (*) & 2269 (*) & 55.6 (*) & 509  (*) \\ \hline
\end{tabular}
\caption{\label{tab_models} Performance of the main generic models for
complex networks. For each network, we give its number
of vertices $n$, its number of links $m$,
its clustering $c$, and
its average distance $d$. Moreover, we give the values of these
parameters for typical graphs with the same number of vertices and
edges obtained with ER, MR, AB and WS models.
Moreover, in The cases pointed by a star (*), the real
clustering is too large to be obtained with the WS model. Therefore we
used in these cases the parameters inducing the maximal clustering,
which yields very large average distances.}
\end{center}
\end{table}

\medskip

Many other attempts have been made to reach the goal of obtaining
models which give graphs having each of the three main properties 
we have cited. Most of them are described in
\cite{albert02statistical,dorogovtsev02evolution,strogatz01exploring}.
Some deterministic models, which we do not detail here, have also been introduced
\cite{barabasi01deterministic,comellas03vertex} which produce the wanted
properties and are suitable for analysis. However, they cannot be
considered as realistic and the obtained graphs have specific
properties which make them very different from real-world \cns.

Another model is the Dorogovstev and Mendes (DM) model which generates
highly clusterised \networks\ with a power law degree distribution
\cite{dorogovtsev02evolution}. In this model, at each step a \node\
is created, an \link\ is chosen at random and the new \node\ is
connected to both extremities of the \link. Since high-degree \nodes\
have more \links, they are more likely to be chosen. The hidden
preferential attachment and the creation of triangles induces the
wanted properties. However the parameters of this model cannot be
tuned and it has some unexpected properties (for instance, there is no
node of degree $1$ and it produces {\em planar} graphs\,\footnote{A
graph is planar iff it can be embedded in the plane so that no edges
intersect. Since \links\ are transformed in triangles, one can easily
be convinced that one step of the algorithm does not create \links\
intersections.}). Therefore we are not going to use it hereafter.

\medskip

More recent models
\cite{barrat05rate,holme02growing,serrano05tuning,serrano05competition,vazquez03growing,vazquez03modeling,wang05evolving}
have been introduced. Some of them use the preferential
attachement principle, and create some triangles at each step. In
\cite{holme02growing}, for each newly created \link, a triangle is
built using both extremities of this \link\ and a neighbor of the
old \node. In \cite{wang05evolving}, a step of the algorithm consists
either in the creation of a new \node\ with $m$ \links\ or in the
creation of $m$ \links\ between neighbors of preferentially chosen
\nodes.

The model introduced in \cite{serrano05tuning} is an evolution of the
configuration model in which both the degree distribution and the
clustering distribution can be pre-defined. Connections points are
created for each \node\ in the future network and a first step
consists in creating triangles using these connections points. A
second step closes the \network\ by connecting unused connections.

\bigskip

In this paper, we propose a solution to the random sampling of graphs
which have all the three wanted properties. To achieve this, we focus
on another property of {\em all} real-world \cns, namely their
underlying bipartite structure, in Section~\ref{sec_bipartite}. We then
propose two models: the random sampling of bipartite graphs with
prescribed degree distributions, and the growing bipartite model with
preferential attachment in Section~\ref{sec_models}.  Indeed, as shown
in Sections~\ref{sec_analysis} and \ref{sec_experimental},
respectively formally and experimentally, these models induce the
three wanted properties. This means that they can be viewed as
consequences of the underlying bipartite structure of all \cns, which
is our main contribution.

\section{\label{sec_bipartite}Complex networks as bipartite graphs}

A bipartite \network\ is a triple $G=(\top,\bot,E)$ where $\top$ and
$\bot$ are two disjoint sets of \nodes, respectively the top and
bottom \nodes, and $E\subseteq \top\times \bot$ is the set of \links.
The difference with classical \networks\
lies in the fact that \links\ exist only between top \nodes\ and
bottom \nodes.

Two degree distributions can naturally be associated to such a
\network, namely the {\em top degree distribution}: $\top_k = \frac{|
\{ t \in \top: d(t)=k \} |}{|\top|}$ and the {\em bottom degree
distribution}: $\bot_k = \frac{| \{ t \in \bot: d(t)=k \}
|}{|\bot|}$.  These two distributions play a central role in the
following.

\subsection*{Natural bipartite structures}

As already noticed for instance in
\cite{ipl04,newman01arbitrary}, some
\cns\ display a natural bipartite structure. Among our examples, one
can view \Actors\ (two actors are linked if they are part of a same
cast) as a bipartite \network\ where $\top$ is the set of movies, $\bot$
is the set of actors, and each actor is linked to the movies he/she
played in.  \Coauthoring\ can also be viewed this way with $\top$
being the set of papers and $\bot$ being the set of authors, each
author being linked to the papers he/she (co-)signed. Likewise, in
\Cooccurrence\ one can link each sentence to the words it contains.

\begin{figure}[!h]
\begin{center}
\myincludegraphics[scale=0.6]{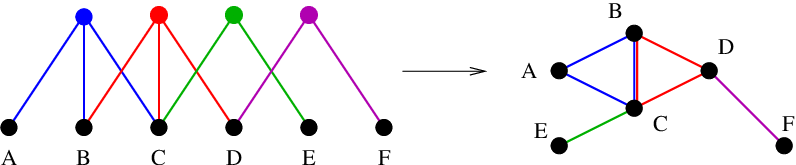}
\caption{\label{bipartitevision}A bipartite network and its $\bot$-projection.
Notice that the link $\{B,C\}$ is obtained twice since $B$
and $C$ have two neighbors in common in the bipartite network.}
\end{center}
\end{figure}

Given a bipartite \network\ $G=(\top,\bot,E)$, one can easily obtain
its classical version, also called the one-mode or $\bot$-projection,
defined as $G'=(\bot,E')$ where $\{u,v\}$ is in $E'$ if $u$ and $v$
are both  connected to a same (top) node in $G$. See
Figure~\ref{bipartitevision} for an example. From the bipartite
versions of \Actors, \Coauthoring\ and \Cooccurrence\ \networks, one
can then recover their classical versions. In the $\bot$-projection of
a bipartite \network, each top \node\ induces a clique (complete
subgraph) between the bottom \nodes\ to which it is linked: all actors
of a given movie have played together therefore they must be all
linked.

However, given the $\bot$-projection of a bipartite \network, it is in
general not possible to recover the bipartite \network\ from which it
has been obtained in an unique way. Similarly if a \network\ is not
naturally bipartite there may exist many bipartite versions of it.

\subsection*{Recovering a bipartite structure}

For the sake of completeness, we now recall and detail
the decomposition scheme we proposed in \cite{ipl04} which, given a
\network, produces a bipartite \network\ whose $\bot$-projection is
the initial \network. The aim of this scheme is to obtain a bipartite
graph whose properties are similar to the ones met in natural
bipartite graphs, namely the number of top \nodes\ has the same order
of magnitude as the number of bottom \nodes\ and there are some
high-degree top nodes (see below and Figure~\ref{fig_distr_top_bot}). 

First notice that the decomposition scheme is nothing but a clique
covering problem: it computes a set of cliques (which will correspond
to the top nodes in the bipartite graph) such that each \link\ belongs
to at least one clique (which ensures that the  $\bot$-projection of
the decomposition is exactly the original \network). Simple ideas to
cover the graph with cliques might be to consider each edge as a
clique, or to consider all maximal cliques. However, the first
approach would not yield large cliques while the second one could
yield too many cliques (the number of maximal cliques may be
exponential).

To reach our goal, we proposed \cite{ipl04}
the following decomposition. We pick
for each edge a largest clique containing it: a clique whose size is
maximal among the ones containing the edge. Notice that this clique
may contain only two \nodes. Moreover, if there are several such
cliques for the same \link, we pick one at random. This decomposition
ensures the complete covering of the \network. Moreover, the number of
cliques is at most equal to the number of \links, which is of the
order of the number of \nodes. Finally, since we take largest cliques,
we expect to find most of the large cliques contained in the \network.

In the case of Figure~\ref{bipartitevision} we obtain several cliques
of size 2 (namely $\{C,E\}$ and $\{D,F\}$), and we have to choose at
random between $\{A,B,C\}$ and $\{B,C,D\}$ when considering the \link\
$\{B,C\}$.  However, these two cliques are obtained from other \links,
and we finally obtain a unique decomposition which is nothing but the
bipartite \network\ on the left of the figure.

\medskip

\begin{figure}[!h]
\begin{center}
\myincludegraphics[scale=0.37]{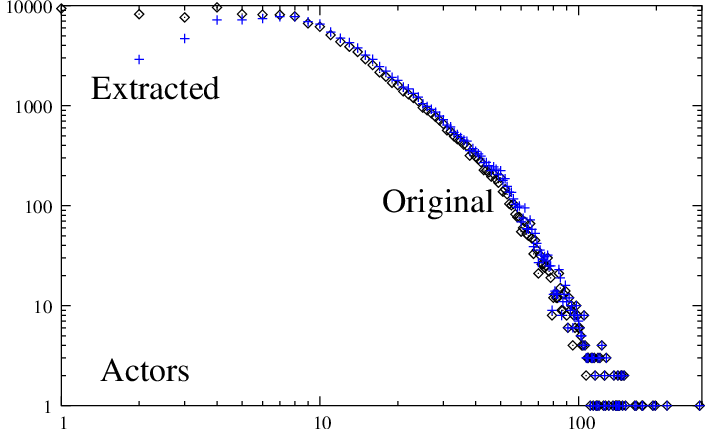}
\myincludegraphics[scale=0.37]{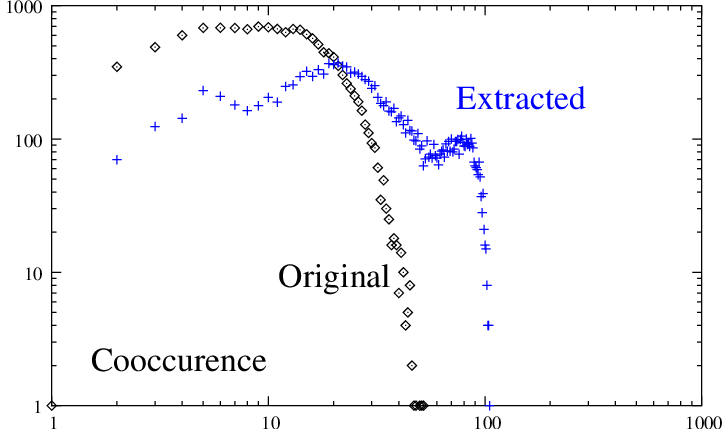}
\myincludegraphics[scale=0.38]{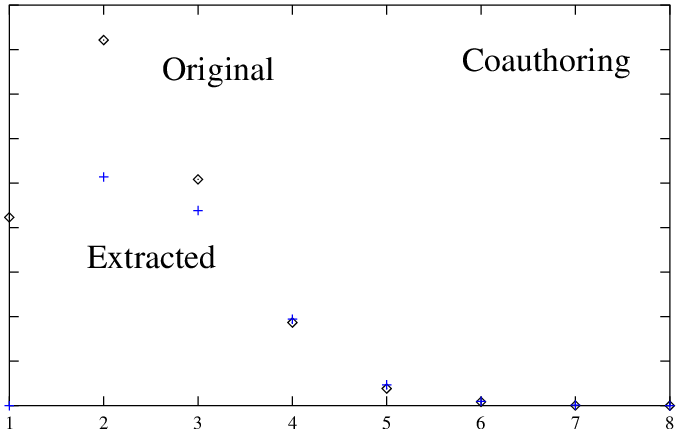}
\caption{\label{comparecliquesizedist}Original clique size
distribution for \Actors, \Cooccurrence\ and \Coauthoring, and
extracted clique sizes distribution with the decomposition scheme.}
\end{center}
\end{figure}

The central aim of our decomposition scheme is, given the
$\bot$-projection of a natural bipartite \network, to produce an
artificial bipartite \network\ similar to the original bipartite
\network\ itself. A way to evaluate it is therefore to decompose the
$\bot$-projection version of a natural bipartite \cn\ and to compare
the obtained bipartite network to the original one. This is what we do
in Figure~\ref{comparecliquesizedist}.

The obtained distributions display some differences for the three
\networks\ decomposed. First, the decomposition scheme produces no
cliques of size $1$ since the smallest extracted element is the \link\
(a $2$-clique). Moreover, many $2$-cliques have not been found, which
means that these $2$-cliques are not maximal in the original
graph. For cliques of size more than $2$, our extraction algorithm has
been able to find most cliques, or even more. Such new large cliques
are induced by the overlapping of other cliques. Notice that in the
case of \Cooccurrence\ there are many new very large cliques which
have been created by overlapping, while in \Actors\ and \Coauthoring\
this phenomena is very weak. We will discuss more about the
overlapping in the conclusion. Despite these differences, the obtained
size distributions are similar to the original ones in the sense that
we obtain a nontrivial number of large cliques and a similar number of
cliques, which are the two main points here.

\subsection*{Practical computation}

\begin{figure}[!h]
\begin{center}
\myincludegraphics[scale=0.5]{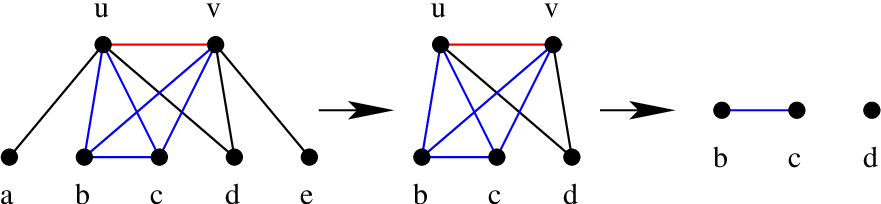}
\caption{\label{cliqueinneigh}Given a graph $G=(V,E')$, we are looking for
a largest clique containing the edge $\{u,v\}$. This clique is
necessarily contained in the subgraph induced by
$N(u) \cup N(v)\cup \{u,v\}=\{b,c,d,u,v\}$. It is actually sufficient to
compute the largest clique $\mathcal{C}$ in the subgraph induced
by $N(u)\cap N(v)=N(u,v)=\{b,c,d\}$ since the clique we are looking for is nothing
but $\mathcal{C}\cup \{u,v\}$ which, in our case, gives $\{u,v,b,c\}$}
\end{center}
\end{figure}

Notice that minimizing the number of cliques leads to the {\em minimal
clique covering problem} which is known to be NP-complete
\cite{monson95survey,orlin77contentment}. Computing maximal cliques of
a \network\ is also NP-complete \cite{abello99maximum,bomze99maximum}
and so is the computation of the largest clique containing a given
\link\ $\{u,v\}$. However, some heuristics make it possible to compute
it if the \network\ is not too large. In our case, we use the
following remarks. Let us denote the sets of neighbors of a \node\ and
an \link\ by $N(u)=\{v\in V | \{u,v\}\in E\}$ and $N(u,v)=N(u)\cap
N(v)$ respectively. The largest clique containing
$\{u,v\}$ in $G$ is $\mathcal{C}\cup \{u,v\}$, where $\mathcal{C}$ is
the largest clique in the sub-\network\ of $G$ induced by
$N(u,v)$. Figure~\ref{cliqueinneigh} illustrates this process.

\medskip

\begin{figure}[!h]
\begin{center}
\begin{tabular}{ccc}
\myincludegraphics[scale=0.3]{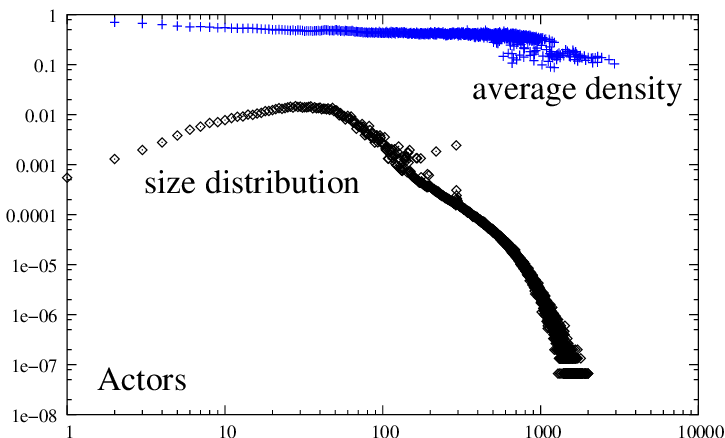} &
\myincludegraphics[scale=0.3]{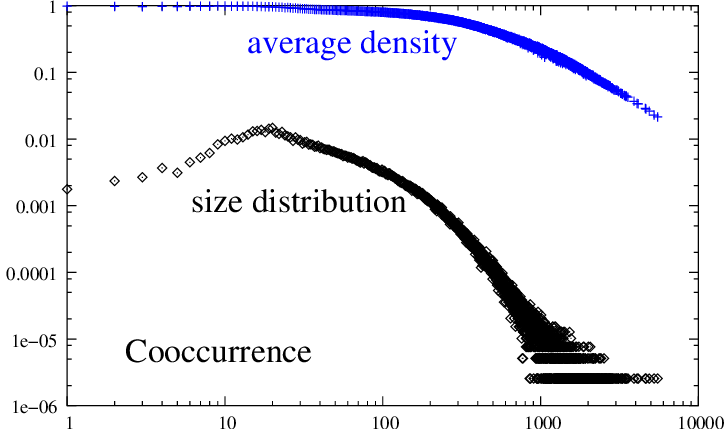} &
\myincludegraphics[scale=0.3]{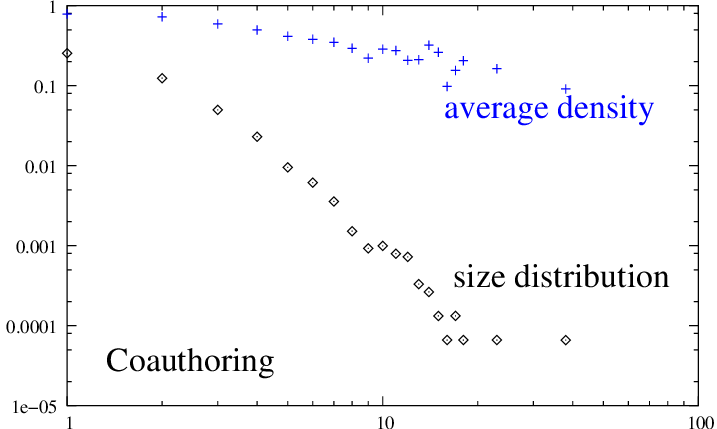} \\
\myincludegraphics[scale=0.3]{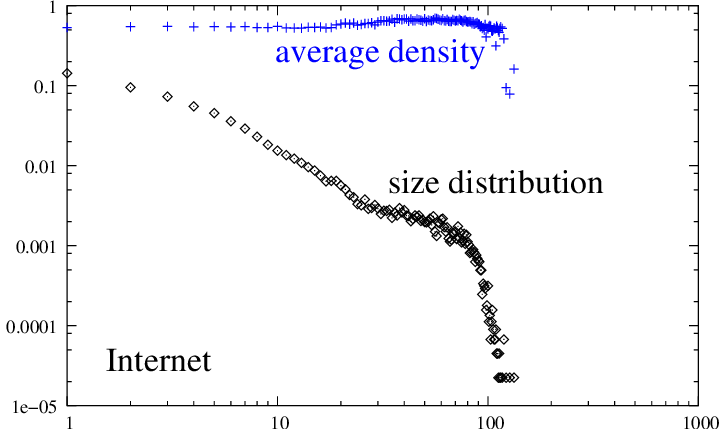} &
\myincludegraphics[scale=0.3]{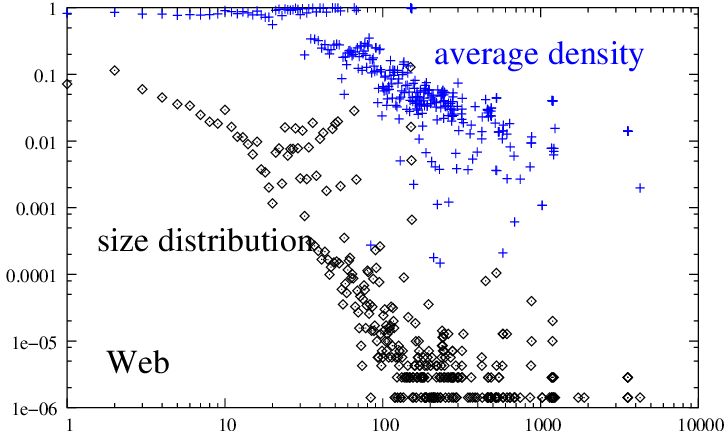} &
\myincludegraphics[scale=0.3]{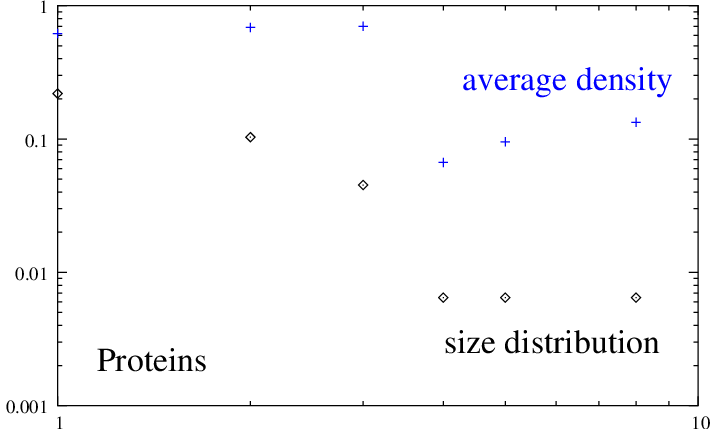}
\end{tabular}
\caption{\label{neighsize}
Distribution of the $N(u,v)$ sizes for all edges $(u,v)$, and average
density of neighborhoods of given size.}
\end{center}
\end{figure}

Recall that the decomposition process relies on a NP-complete problem in
general. However, we observed that in real-world \cns, the
sub-\networks\ induced by $N(u,v)$ for all \links\ $\{u,v\}$ are in
general very dense and very small (Figure~\ref{neighsize}), which is
due to the high clustering and to the power law degree distribution,
respectively. The small size makes it possible to compute the largest
clique containing $\{u,v\}$ very efficiently in practice.

\subsection*{Properties of the bipartite graphs}

\begin{figure}[!h]
\begin{center}
\begin{tabular}{ccc}
\myincludegraphics[scale=0.3]{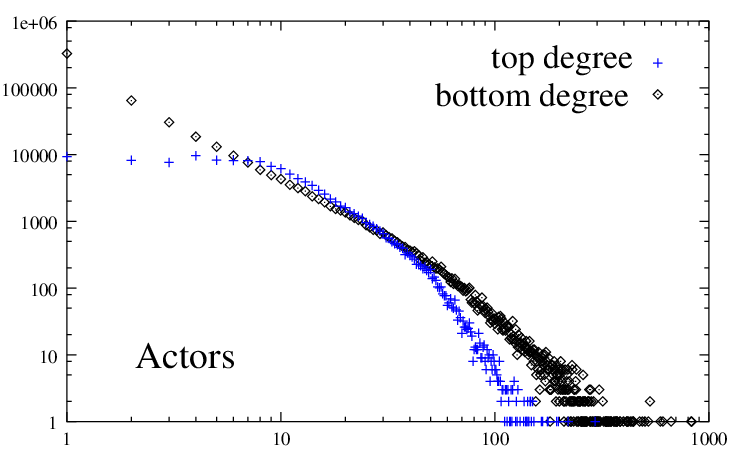} &
\myincludegraphics[scale=0.3]{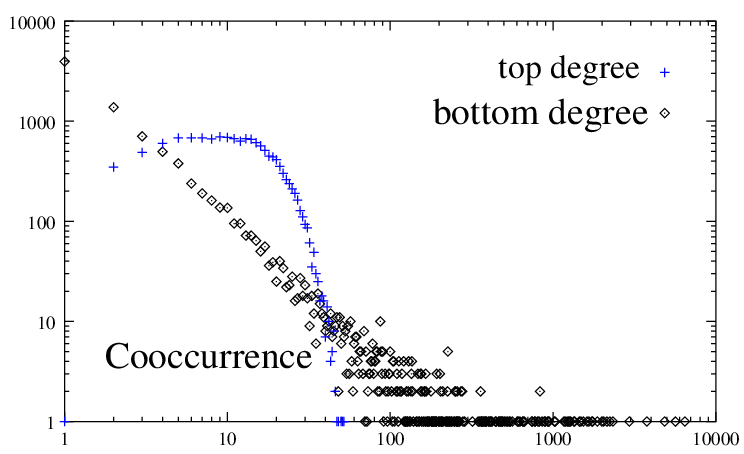} &
\myincludegraphics[scale=0.3]{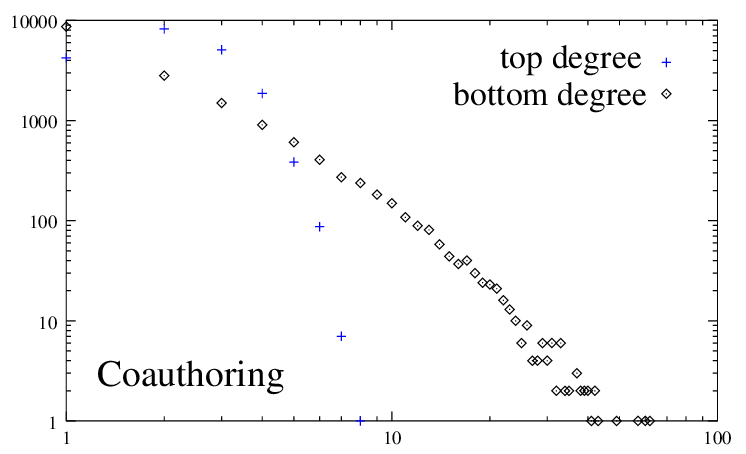} \\
\myincludegraphics[scale=0.3]{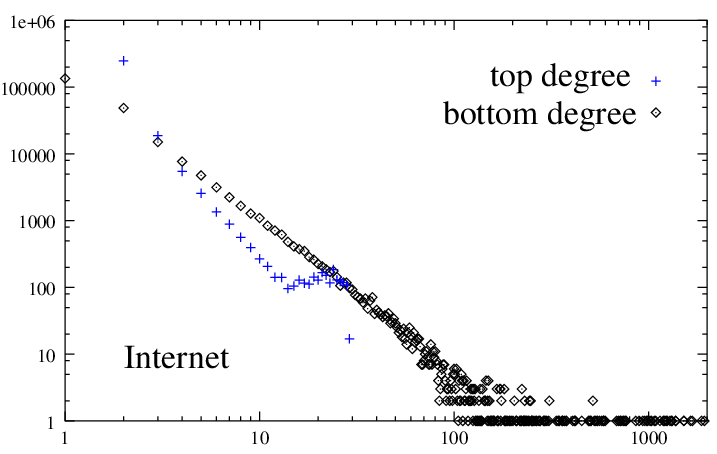} &
\myincludegraphics[scale=0.3]{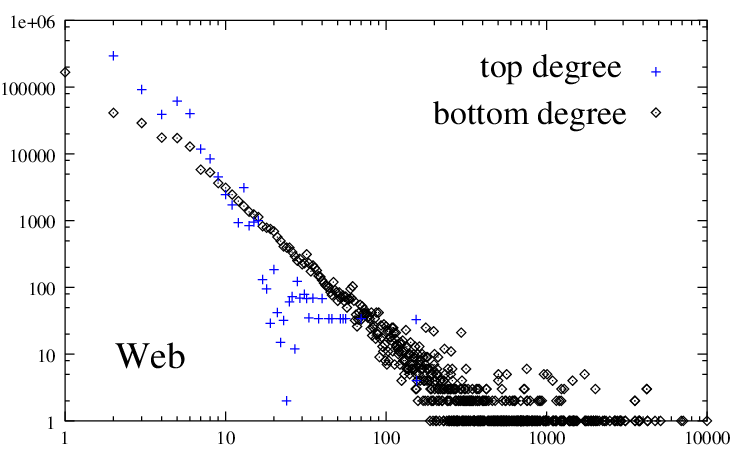} &
\myincludegraphics[scale=0.3]{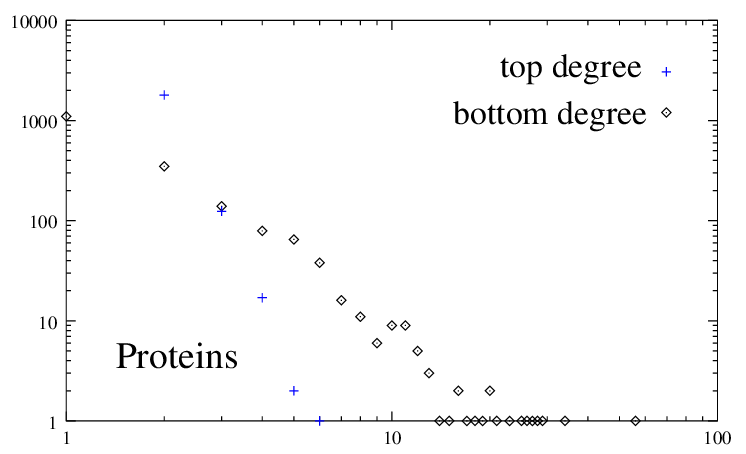}
\end{tabular}
\caption{\label{fig_distr_top_bot}
Top and bottom degree distributions for the natural bipartite versions
of \Actors,
\Cooccurrence, and \Coauthoring, and for the bipartite version of
\Internet, \Web, and \Proteins\ obtained with the decomposition scheme.}
\end{center}
\end{figure}

Given the general decomposition scheme, we can now transform
any \cn\ into a bipartite \network. Figure~\ref{fig_distr_top_bot}
shows the top and bottom degree distribution for the natural
bipartite networks \Actors, \Cooccurrence\ and \Coauthoring, and
the ones obtained for \Internet, \Web\ and \Proteins\ \networks\ 
using our decomposition scheme.

All these distributions have a property in common: bottom degree
distributions fit very well power laws in all cases. On the contrary,
the top degree distributions are of two kinds: while \Cooccurrence,
\Coauthoring, \Internet\ and \Proteins\ ones exhibit a Poisson
behavior, \Actors\ and \Web\ ones are more heavy tailed.

These results lead to several remarks. First, the presence of a power
law bottom degree distribution seems universal, just like the power law
distribution in the $\bot$-projection of these graphs, but the
effective exponent is not and depends on the considered \cn. Second,
the top degree distributions can be qualitatively different and this
point is important in the use of the bipartite structure for modeling
\cns\  since it can impact on some characteristics of the generated
\networks. Further remarks will be pointed out in
Section~\ref{sec_conclusion}.

\begin{figure}[!h]
\begin{center}
\begin{tabular}{ccc}
\myincludegraphics[scale=0.3]{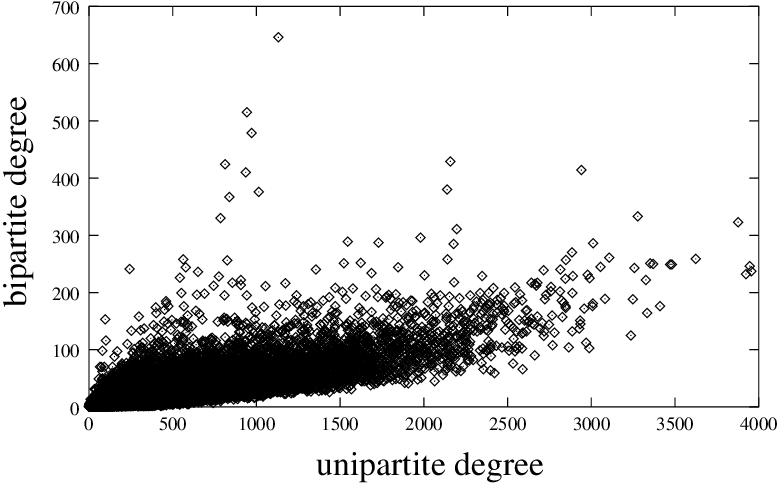} &
\myincludegraphics[scale=0.3]{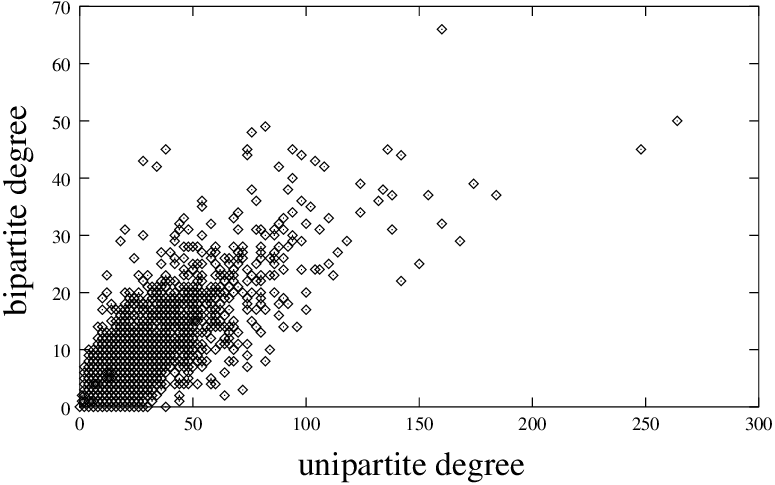} &
\myincludegraphics[scale=0.3]{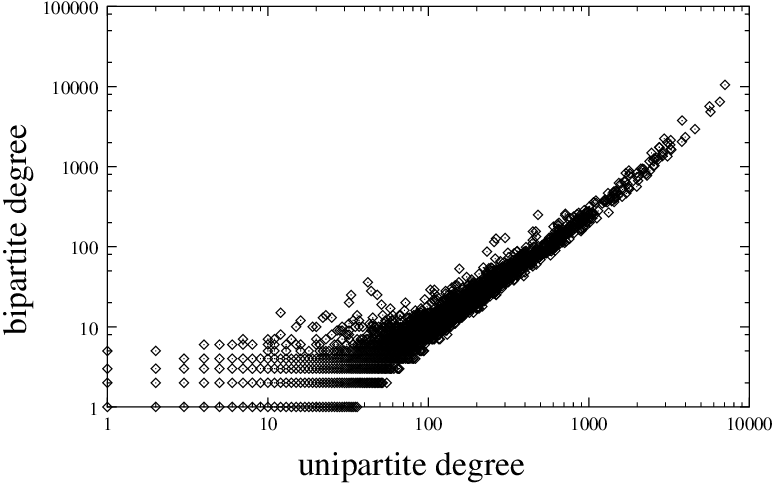} \\
\end{tabular}
\caption{\label{degreecorrelation}Correlations between degrees in the
classical network and the bipartite one. From left to right: \Actors,
\Coauthoring, and \Cooccurrence.}
\end{center}
\end{figure}

Note also that the degree of a \node\ in the $\bot$-projection is the
sum of the degrees of the top \nodes\ to which it is 
connected in the bipartite graph, minus the number of vertices in
common in the neighborhood of these \nodes. One can easily be
convinced that this overlap between neighborhoods, if any, can have a
great impact on the degree distribution. To deepen this notion of
overlap, one can observe the correlation between the bottom \node\
degrees in both bipartite and $\bot$-projection
(Figure~\ref{degreecorrelation}). There exists nontrivial correlations
in both cases, which are particularily strong in the case of
\Cooccurrence. Others remarks on the overlap will be discussed further
in Section~\ref{sec_conclusion}.

\bigskip

Finally, we have shown in this section that {\em all} \cns\ have a
nontrivial underlying bipartite structure, which can be computed using
our decomposition scheme. This leads us to the following question: is it
possible to see the main properties of real-world \cns\ as
consequences of their underlying bipartite structure? We answer this
question in the next sections.

\section{\label{sec_models}The bipartite models.}

Our aim is now to use the new general property of real-world
\cns\ discovered in the previous section, namely their
underlying bipartite structure, as a way to propose a model
which captures the main wanted properties.

As discussed in the first section of this paper, there are
basically two ways to achieve this goal. First, we may try
to sample random bipartite graphs with prescribed (top and
bottom) degree distributions. Second, we may try to propose
a construction process similar to the ones observed in practice,
to obtain a {\em growing} model.

We proposed such models  in \cite{ipl04}. In order to deepen the
understanding of these models, we here recall and discuss more
precisely their definitions,
and we provide a full (both analytic and  experimental) analysis in
the next sections.

\subsection*{Random sampling of bipartite \networks\ with prescribed
degree distribution}

\begin{figure}[!h]
\begin{center}
\myincludegraphics[scale=0.6]{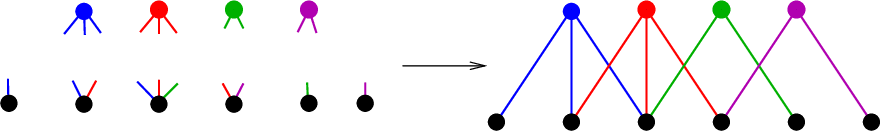}
\caption{\label{modelebiparti}Construction of a random bipartite
graph with prescribed degree distributions: first top and bottom
vertices are drawn and each vertex is assigned a degree with respect to the
given distributions, then edges are chosen randomly between the two
sets.}
\end{center}
\end{figure}

One can sample uniformly a random
bipartite \network\ with prescribed (top and bottom) degree
distributions in the spirit of the configuration model as follows (see
Figure~\ref{modelebiparti}) \cite{progurl,newman01arbitrary,newman01random}:
\begin{enumerate}
\item generate both top and bottom \nodes\ and assign to each \node\ a
degree drawn from the given distributions,
\item create for each \node\ as many connection points as its degree,
\item link top and bottom connection points randomly,
\end{enumerate}

This process generates random bipartite \networks\ uniformly within
the set of bipartite \networks\ with the given degree
distributions. However it cannot be used without taking care of the
following constraints: top and bottom distributions cannot be arbitrary
since they must allow the total degree of both sets to be
equal. Actually one only has to ensure that the number of \nodes\ times 
the mean of the distribution brings the same value for top and bottom
sets. The second problem arises with the fact that even if the two
distributions are theoretically consistent, two sets of degrees
experimentally drawn from these distributions can be inconsistent (the
sums of the degrees are different). A classical trick, which induces no bias,
consists in
dropping one top and one bottom \node\ at random and redraw their
degree \cite{newman01arbitrary,newman01random}. This last step may
have to be done more than once before one obtains correct values but
finally the implementation and use of the model is very simple and
efficient \cite{progurl}.

Note that, just like with the MR model, multiple \links\ may appear.
Again, one can easily show that they can be neglected when the graph is
large. Moreover, some approaches exist
\cite{latapy05random,milo03uniform} which can easily be modified to obtain
random bipartite \networks\ without multiples \links. This is however
out of the scope of this paper.

\subsection*{Growing bipartite model with preferential attachment}

The random bipartite model assumes that two distributions, for both
top and bottom degrees, are explicitly given. One can also use other
rules (preferential attachment for instance) to define them implicitly
and introduce a growing model. Indeed, as already noticed, the bottom
degree distributions follow a power law. This leads to the following
model: at each step, a new top \node\ is added and its degree $d$ is
sampled from a prescribed (top) distribution (which qualitatively
varies between \networks). Then, for each of the $d$ \links\
of the new \node, either a new bottom \node\ is added (with
probability $1-\lambda$) or one is picked among the preexisting ones
using preferential attachment (with probability $\lambda$). The
parameter $\lambda$ is the {\em overlap ratio}, defined as the average
ratio of preexisting bottom \nodes\ to which a new top \node\ is
connected.

It is generally not possible to know exactly the order in which
cliques are created on real-world bipartite \networks, but the average
ratio can be computed globally as $\lambda=1-\frac{|\bot|}{\sum
d_{\top}}$. One can compute it and get $0.733$ for \Actors, $0.877$
for \Coauthoring\ and $0.949$ for \Cooccurrence. Notice that
$1-\lambda$ can be rewritten and is simply the inverse average bottom
degree (since $\sum d_{\top}=\sum d_{\bot}$), therefore a high overlap
ratio yields a high average bottom degree (since only few nodes are
created at each time step).

At each step of the construction process, the bipartite \network\ has
the required degree distributions: the prescribed top degree
distribution is obtained by construction while the power law degree
distribution is obtained using preferential attachment, which can be
shown formally in exactly the same way as in the original AB model
\cite{barabasi99emergence}. Notice moreover that this construction
process is very similar to the one observed in some real-world
cases. For instance, \Actors\ is built exactly this way: when a new
movie is produced (which corresponds to the addition of a top \node),
it is linked to actors according to their popularity, and to some new
actors, playing in a movie for the first time.

\medskip

We finally have two models to produce bipartite networks similar to
the ones obtained from real-world \cns, in terms of top and bottom
degree distributions. The next question is to ask if they capture
the other properties of interest in their $\bot$-projection, namely
the average distance, the degree distribution and the clustering. We
will answer positively to this question with formal arguments and
with experimental results in the next sections.

\section{\label{sec_analysis}Analysis of the models.}

Our aim in this section is to give formal proofs for the main
properties of the $\bot$-projection of a random bipartite graph with
prescribed degree distributions. Some of these properties, and others,
have been studied independently in
\cite{newman01arbitrary,newman01random} with different techniques and
a different point of view. We however believe that our proofs give new
insight on these properties, therefore we give them below. In
particular, our proof techniques may be considered as more
mathematically rigorous.

Since these properties are induced by a {\em typical} graph (this is
what random sampling gives us), this is a way to answer the following
question: what properties are induced by the underlying bipartite
structure? In particular, can we see the main properties of real-world
\cns, namely low average distance, power law degree
distribution and high clustering, as consequences of the underlying
bipartite structure?

We will see that it is indeed the case. Notice that many other
properties, like the size distribution of the connected
components for instance, are of high interest. It is shown in \cite{newman01arbitrary}
that under reasonable conditions on the degree distributions the
$\bot$-projection is connected, or at least has a giant component. In all the practical cases,
these conditions are fulfilled, therefore we will restrict ourselves
to this case.

\medskip
\subsection*{Degree distribution}

Let us first consider the degree distribution of the $\bot$-projection
of a random bipartite \network\ $G=(\top,\bot,E)$. Given a bottom
\node\ $u$, we denote by $d(u)$ the degree of $u$ in the bipartite
\network, and by $d_U(u)$ its degree in its $\bot$-projection. We want
to study the distribution of $d_U(u)$ (we actually deal here with the
expected value for a randomly chosen $u$).

\begin{lemma}
\label{Lemma:setinter}
Let us consider a bottom \node\ $u\in \bot$. The 
expected number of bottom \nodes\
which have a neighbor (in $\top$) in common with $u$, \ie\ $d_U(u)$, is:
\begin{equation*}
\frac{d(u)}{|\top|}\cdot \sum_{t \neq u} d(t) +{\cal O}\left(\frac{d(u)^2}{|\top|^2}\cdot\sum_{t \neq u} d(t)^2\right)
\end{equation*}
\end{lemma}

\medskip

\noindent
\begin{proof}
The exact expected value of $d_U(u)$ is given by
$
d_U(u)=\sum_{t \neq u} \left( 1-\frac{{|\top|-d(u) \choose d(t)}}
                              {{|\top| \choose d(t)}}
                \right)
$
since the probability that a given bottom \node\ $t$ has a top
neighbor in common with $u$ depends only on the degree of both \nodes\
and the number of top \nodes. To simplify this formula, we can
approximate the ratio ${|\top|-d(u) \choose d(t)}/{|\top| \choose
d(t)}$ as follows:
$
\frac{{|\top|-d(u) \choose d(t)}}{{|\top| \choose d(t)}}
= \frac{(|\top|-d(u))!(|\top|-d(t))!}{|\top|!(|\top|-d(u)-d(t))!}
\sim \frac{(|\top|-d(t))^{d(u)}}{|\top|^{d(u)}}
\sim 1-\frac{d(t) d(u)}{|\top|}+{\cal O}\left(\left(\frac{d(t) d(u)}{|\top|}\right)^2\right)
$.

\noindent
Therefore:
$
d_\top(u)
\sim \sum_{t \neq u} \left(\frac{d(t) d(u)}{|\top|}+{\cal O}\left(\left(\frac{d(t) d(u)}{|\top|}\right)^2\right)\right)
\sim \frac{d(u)}{|\top|} \sum_{t \neq u} d(t) +{\cal
O}\left(\frac{d(u)^2}{|\top|^2}\sum_{t \neq u} d(t)^2\right)
$,
which is the formula of the claim.
\qed
\end{proof}

\medskip

This lemma makes it possible to compute the probability for a \node\
$u$ in the $\bot$-projection \network\ to have a given degree $k$ if the
bottom degree distribution is a power law with exponent $\beta$:
$$
P[d_U(u)=k] \sim P[d(u)=\frac{n}{\sum_{t \neq u} d(t)}\cdot k]
	    \sim \frac{1}{(\sum_{t \neq u} d(t))\cdot k)^\beta} \sim k^{-\beta}
$$

Therefore, as long as the bottom degree distribution follows a power law,
the degree distribution in the $\bot$-projection of the \network\
also follows a power law with the same exponent, which is indeed the
case in practice as one can check in Figures~\ref{fig_model_deg_dist}
and~\ref{fig_distr_top_bot}. The main reason comes from the fact that
$\bot$-\nodes\ are somehow randomly linked to $\top$-\nodes\ and their
degree in the projection is therefore strongly correlated with their
bipartire degree.

\medskip

\subsection*{Average distance}

To study the average distance in the $\bot$-projection of a
\network\ obtained with the model, we will use a result from L.~Lu
about the diameter (\ie\ the largest distance between any two \nodes)
of some specific random \networks:

\begin{theorem}\cite{lu01diameter}
\label{Th:lu}
Let $G=(V,E)$ be a \network\ whose \nodes\ are weighted with weights
$w_1,\cdots ,w_n$, such that each \link\ $\{i,j\}$ appears with
probability $w_i\cdot w_j\cdot p$. If the degrees of the \nodes\ in
$V$ follow a power law with an exponent $\beta$ strictly greater than
$2$, then the diameter of the \network\ $G$ is almost surely
$\Theta(\log(n))$\footnote{$f=\Theta(g)$ if and only if $f={\cal O}(g)$ {\em and} $g={\cal O}(f)$i.}.
\end{theorem}

This theorem, together with the one presented above on the degree
distribution of the $\bot$-projection of the \network, leads to the
following result:

\begin{theorem}
Let $G=(\top,\bot,E)$ be a bipartite \network\ such that the bottom degree
distribution follows a power law with an exponent greater than $2$.
Then the diameter of the $\bot$-projection of $G$ is almost
surely $\Theta(\log(|\bot|))$.
\end{theorem}

\begin{proof}
Given two bottom \nodes\ $u$ and $v$ in $\bot$, the probability that they
are connected in the $\bot$-projection is equal to the probability
that they are both linked to a same top \node\ in $G$. This
probability is exactly proportional to $d_\bot(u)\cdot d_\bot(v)$. Therefore we
can apply Theorem~\ref{Th:lu} considering that the weight of each
\node\ is its degree and so the connection probability is ensured, and
as long as bottom degree distribution follows a power law with an
exponent $\beta$ strictly greater than $2$. The diameter of the
$\bot$-projection of the \network\  therefore is almost surely
$\Theta(\log(|\bot|))$.
\qed
\end{proof}

Since the diameter is an upper bound for the average distance,
this theorem implies that
the average distance of the $\bot$-projection scales
at most as fast as the logarithm of its number of nodes. Notice that,
as in the case of random networks
\cite{bollobas85random,cohen02structural,cohen03ultrasmall,dorogovtsev03metric,lu01diameter,newman01arbitrary,newman01random},
the average distance may grow
even slower.

\medskip
\subsection*{Clustering}

Recall that the \cc\ of a \node\ $v$ of degree at least $2$
in a \network\ is the probability that two of its neighbors
are linked \cite{watts98collective}, \ie\ the number of
triangles to which $v$ belongs over the number of connected triples
centered on it:
$c(v)=\frac{|\triangle(v)|}{|\wedge(v)|}$. Then the \cc\ of
the graph is defined as:
$\frac{1}{N}\sum_{v, d(v)>1} c(v)$. We define the \cc\ of a
\node\ restricted to a part of its neighborhood as its \cc\
in the subgraph induced by this part of its neighborhood.

Hereafter we give a lower bound for the \cc\ of a \network\ 
$G'$ which is the $\bot$-projection of a bipartite graph
$G=(\top, \bot, E)$ obtained
using the random bipartite model. We show that, under reasonable
assumptions on the top and bottom degree distributions, it is bounded by
a value independent of the size of the graph. This shows that
the model produces graphs with nontrivial clustering.

\medskip

Before entering in the core of this section, note that an
approximation formula for the \cc\ of such a graph is given in
\cite{newman01arbitrary,newman01random}. Here we give an exact formula
for a {\em lower bound}. Both are interesting since the first one
gives an expected value which is indeed very close to the real value,
while the second one gives a guaranty that the exact value is above
the given quantity. We used this approach because we seek qualitative
results only, and so it is sufficient for us to show that the \cc\
does not tend to $0$ when the size of the graph grows. The lower bound
achieves this goal.

First, the probability for two top \nodes\ to have more
than one bottom \node\ in common in their neighborhood tends to
zero when the size of the graph grows. We therefore consider any
vertex $b$ in the $\bot$-projection of the graph and we suppose that
its neighborhood is composed of a set of disjoint cliques. We will
prove the following: \begin{itemize}
\item the effect of the number of top \nodes\ of degree $2$ to which
 $b$ is connected on its \cc\ is negligible, and
\item the \cc\ of $b$ can be bounded by a value which depends only
 on its degree.
\end{itemize}


\medskip

\begin{lemma}
\label{lemma_deg2}
Let $\top_{>2}$ denote the set of top neighbors
of $b$ in $G$ with degree strictly
greater than $2$, and $\bot_{>2}$ denote the set of bottom neighbors of $\top_{>2}$.
Let $p$ be the fraction of neighbors of $b$ which belong to $\bot_{>2}$,
and $\alpha$ be the \cc\ of $b$ (in $G'$) restricted to $\bot_{>2}$.

Then the \cc\ of $b$ in $G'$ scales as $p^2 \cdot \alpha$.
\end{lemma}

\begin{proof}
The fact that the \cc\ of $b$ restricted to $\bot_{>2}$ is $\alpha$
implies that $|\triangle_{\bot_{>2}}(b)|=\alpha \cdot {p\cdot d \choose 2}$.
If we consider the whole neighborhood of $b$, instead of just
$\bot_{>2}$, the number of triangles does not change while the number
of connected triples increases:
$
c(b) = \frac{\alpha \cdot {p\cdot d \choose 2}}{{d \choose 2}}
     = \alpha \cdot \frac{p\cdot d((p\cdot d-1)}{d(d-1)}
     \sim p^2 \cdot \alpha
$
which is the formula of the claim.
\qed
\end{proof}

Therefore, as long as $p$ is a constant, one can neglect the top \nodes\
of degree $2$ when computing the \cc\ of a given \node. Now let us
prove that the \cc\ of a bottom \node\ can be related to its degree.


\medskip

\begin{lemma}
\label{lemma_bound}
If $b$ is connected only to top \nodes\ of degree at least $3$ in $G$,
then: 
$$c(b) \geq \frac{1}{2 \cdot d(b) - 1}$$
\end{lemma}

\begin{proof}
Suppose that $b$ is connected to two top \nodes, $t_1$ and
$t_2$, of degree at least $3$ (we deal with the general case
below).
Then the \cc\ of $b$ is
$c(b) = \frac{{d(t_1)-1 \choose 2} + {d(t_2)-1 \choose 2}}{{d(t_1)+d(t_2)-2 \choose 2}}$
Suppose now that $b$ is connected to $t_2$ and $t_1'$ such that
$d(t_1')=d(t_1)+1$, then the \cc\ of $b$ is
$c'(b) = \frac{{d(t_1)+1-1 \choose 2} + {d(t_2)-1 \choose 2}}{{d(t_1)+d(t_2)-1 \choose 2}}$
and
$
c'(b)-c(b) =\frac{2\cdot(d(t_2)-1)}{(d(t_1)+d(t_2)-2)\cdot(d(t_1)+d(t_2)-3)} 
           > 0$,
which means that the \cc\ grows with the degree of $t_1$ and
$t_2$. A lower bound for the \cc\ of $b$ can therefore be obtained
when both $t_1$ and $t_2$ have the smallest possible degree:~$3$. 

This can be extended to the case where $b$ has more than two top
neighbors to obtain the following lower bound
$
c(b) = \frac{\sum_{t_i}{d(t_i)-1 \choose 2}}{{\sum_{t_i} (d(t_i)-1) \choose 2}}
     \geq \frac{\sum_{t_i}{3-1 \choose 2}}{{\sum_{t_i} (3-1) \choose 2}} \geq \frac{1}{2 \cdot d(b) - 1}
$,
which is the formula of the claim.
\qed
\end{proof}


\medskip

\noindent
The \cc\ of the $\bot$-projection $G'$ can now be easily approximated:
$$c(G') \sim \frac{1}{N}\sum_{b\in\bot}\frac{1}{2 d(b) - 1}$$

As long as there is a linear number $c\cdot N$ of \nodes\ $b$
of degree $2$, the sum scales linearly with $N$:
$\sum_{b\in\bot}\frac{1}{2\cdot d(b) - 1}\geq \sum_{b,
d(b)=2}\left(\frac{1}{2\cdot 2 - 1}\right)= \frac{c\cdot N}{3}$ (we
could have considered \nodes\ of any constant degree $k$ instead of
$2$). Therefore the lower bound for the \cc\ is independent of $N$.
This holds in particular for power law networks since
the number of \nodes\ of degree $2$ is of the order of $N\cdot 2^{-\alpha}$.

Since we do not consider top \nodes\ of degree $2$ in
the last formula (due to Lemma \ref{lemma_deg2}), we must also ensure
that the number of such top neighbors represent at most a constant
fraction (not tending to $1$) of the neighbors. This is indeed the case for
most distributions and in particular for the ones met in practice.
We finally obtain that the \cc\ of the graph is larger than
a non-zero constant independently of the size of the graph. The idea
behind the proof is that each node belongs to a small number of
cliques and even if these cliques are disjoint, the clustering is
still high.

\section{\label{sec_experimental}Experimental results}

The formal results of the previous section give a precise intuition
on how the random bipartite graph model with prescribed degree
distributions behaves. We can also check its properties experimentally
by generating graphs using this model and the same parameters
as the ones measured on real-world \cns. This is what we do in
this section with our six examples, for the purely random
bipartite model as well as for the one with preferential attachment.

More precisely, the networks are generated using the models presented
in section~\ref{sec_models} with $\top$ and $\bot$ distributions
obtained either from the real \networks\ when they are known or from
the decomposition scheme introduced in section~\ref{sec_bipartite}.

\medskip

\begin{figure}[!h]
\begin{center}
\begin{tabular}{ccc}
\myincludegraphics[scale=0.3]{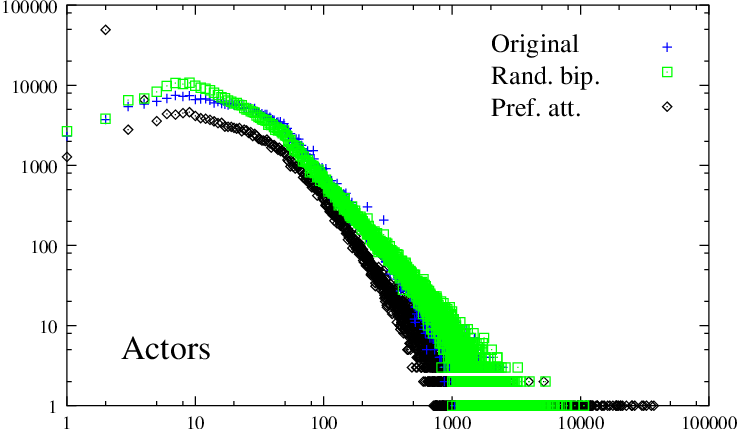} &
\myincludegraphics[scale=0.3]{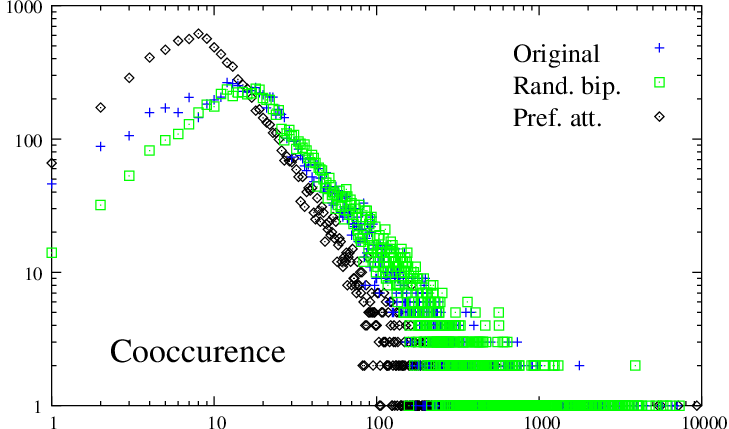} &
\myincludegraphics[scale=0.3]{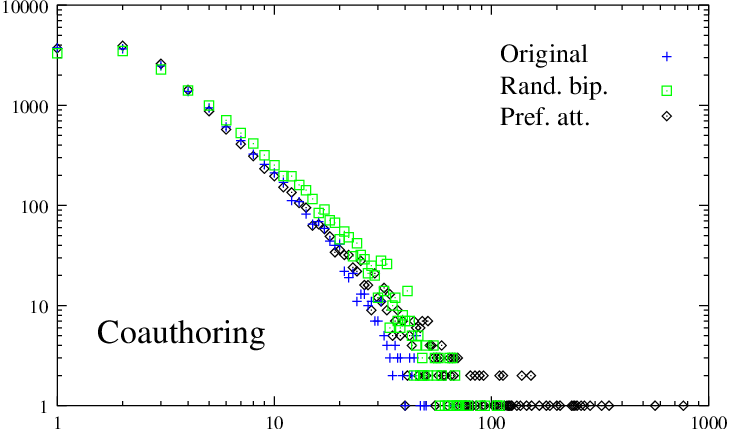} \\
\myincludegraphics[scale=0.3]{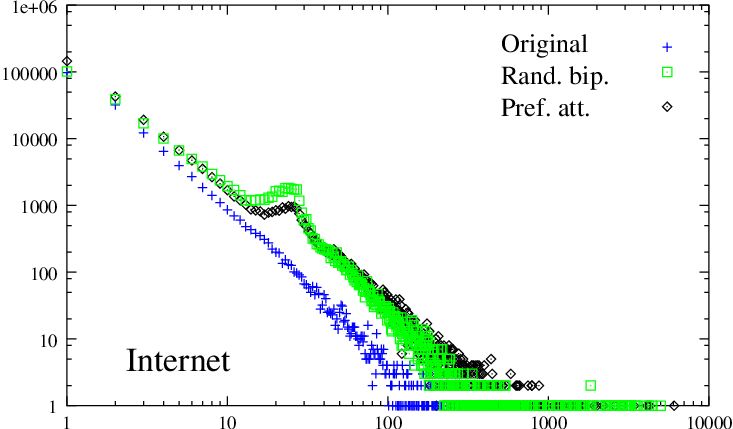} &
\myincludegraphics[scale=0.3]{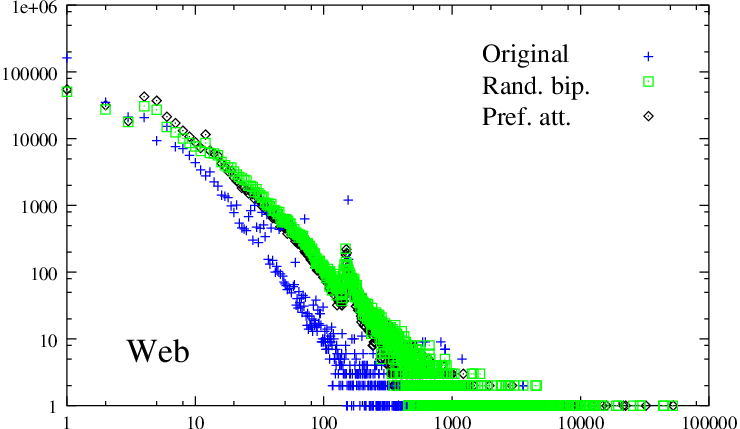} &
\myincludegraphics[scale=0.3]{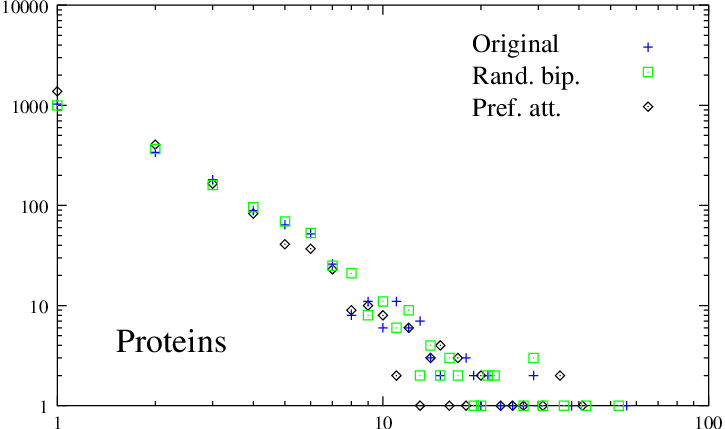}
\end{tabular}
\caption{\label{fig_model_deg_dist}The original degree distribution of
our six examples, together with the ones obtained with the random
bipartite model and with the growing bipartite model.}
\end{center}
\end{figure}

Table~\ref{tab_model_dist} and~\ref{tab_model_cc} give
the values obtained for the average distance and the
clustering. Figure~\ref{fig_model_deg_dist} shows a comparison between
the degree distributions of the original graphs, and the ones obtained
with the two bipartite models.

\begin{table}[!h]
\begin{center}
\small
\begin{tabular}{|l|c|c|c|c|c|c|c|} \hline
       & Internet  & Web     & Actors   & Co-auth  & Co-occur & Protein  \\ \hline \hline
$d$      & 5.80    & 7       & 3.6      & 7.18     & 2.13     & 6.74     \\ \hline
$d_{rb}$ & 2.97    & 3.2     & 3.06     & 5.07     & 2.06     & 5.8      \\ \hline
$d_{gb}$ & 2.81    & 3.53    & 2.83     & 3.98     & 2.6      & 5.45     \\ \hline
\end{tabular}
\caption{\label{tab_model_dist} \small
Average distance of the commonly used models and the bipartite
models. For each network, we give its actual average distance, and the
one obtained with 
the random bipartite model with prescribed degree
distributions $d_{rb}$, and the growing one with preferential
attachment $d_{gb}$.
}
\end{center}
\end{table}

\begin{table}[!h]
\begin{center}
\small
\begin{tabular}{|l|c|c|c|c|c|c|c|} \hline
       & Internet  & Web     & Actors   & Co-auth  & Co-occur & Protein  \\ \hline \hline
$c$      & 0.171   & 0.466   & 0.785    & 0.638    & 0.822    & 0.153    \\ \hline
$c_{rb}$ & 0.32    & 0.663   & 0.767    & 0.542    & 0.831    & 0.187    \\ \hline
$c_{gb}$ & 0.65    & 0.708   & 0.793    & 0.632    & 0.768    & 0.244    \\ \hline \hline
\end{tabular}
\caption{\label{tab_model_cc} \small
Clustering obtained with the commonly used models and the bipartite
models. For each network, we give its actual clustering and the
clustering obtained with 
the random bipartite model with prescribed degree distributions
$c_{rb}$, and the growing one with preferential attachment $c_{gb}$.
}
\end{center}
\end{table}

As expected from the previous section, the graphs obtained with the
random bipartite model have a power law distribution of degrees, a
small average distance and a high clustering. Moreover, by definition,
they have the same distribution of cliques size as the original
network. Therefore the model is qualitatively accurate for the
modeling of general real-world \cns: the simulations fit real-world
values qualitatively well for both \cc\ and average  distance, which
proves the relevance of the underlying bipartite structure as an
essential property to characterize real-world \cns.

\begin{figure}[!h]
\begin{center}
\begin{tabular}{ccc}
\myincludegraphics[scale=0.3]{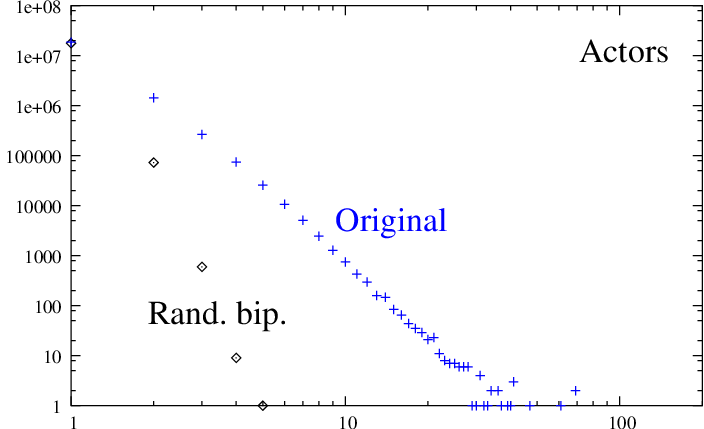} &
\myincludegraphics[scale=0.3]{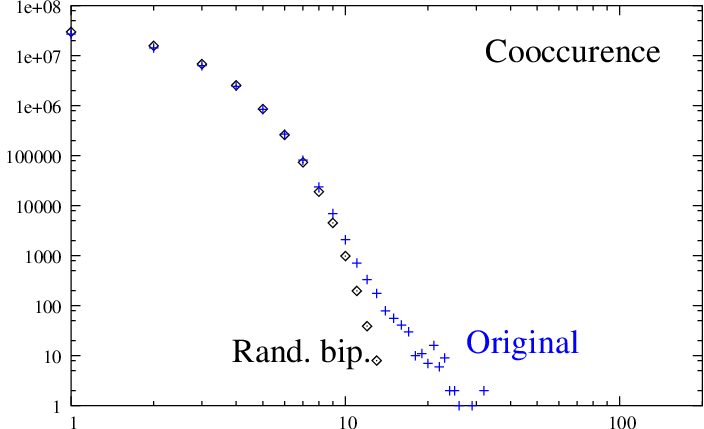} &
\myincludegraphics[scale=0.3]{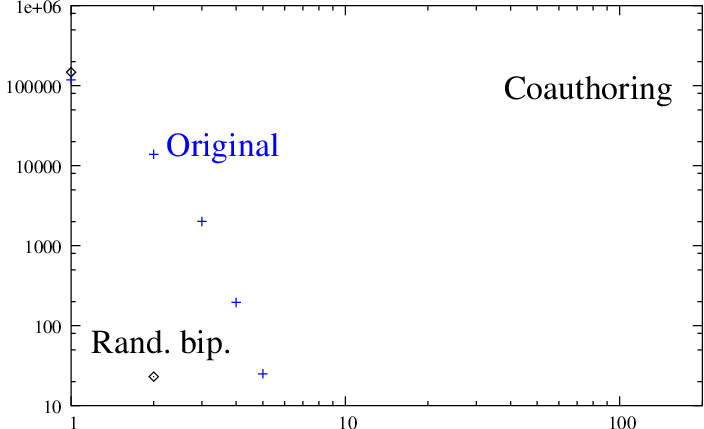} \\
\myincludegraphics[scale=0.3]{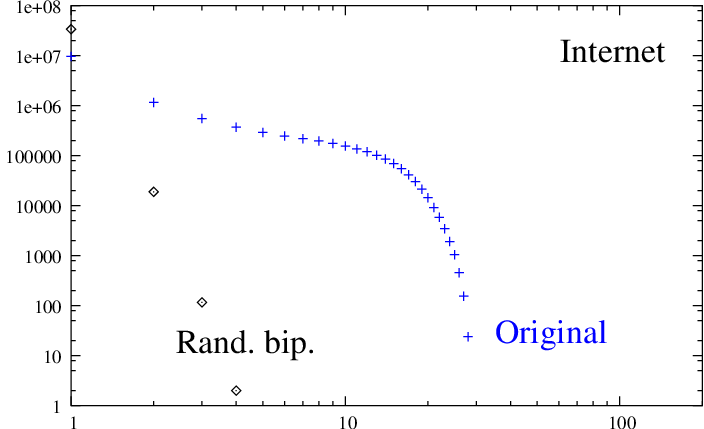} &
\myincludegraphics[scale=0.3]{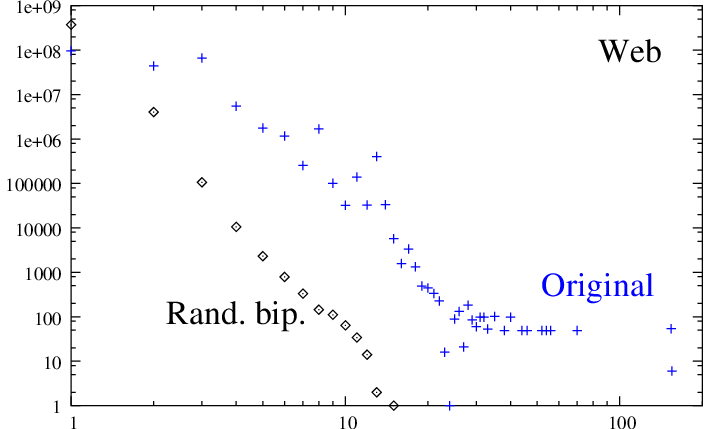} &
\myincludegraphics[scale=0.3]{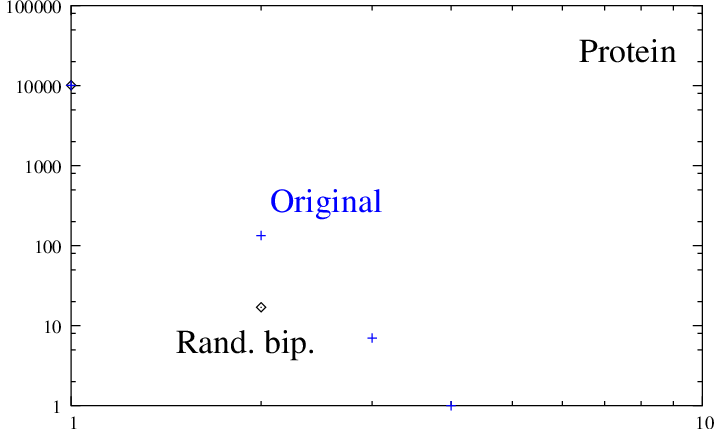}
\end{tabular}
\caption{\label{fig_bip_clustering}Distribution of the size of the
  overlap between cliques (\ie\ intersections of neighbors of top nodes)
  for both original networks (or decomposed
  ones) and random bipartite ones.}
\end{center}
\end{figure}

There are however differences between the values obtained from the
bipartite models and real-world networks. They are consequences of the
following fact: in the original bipartite networks (both natural ones
and the ones obtained from the decomposition), many top nodes have a
large neighborhood intersection. In other words, the overlap between
cliques is large (if two cliques have one neighbor in common, they
certainly have many). This behavior can be viewed as a {\em bipartite
clustering} and is not captured by the bipartite models. The random
linking implies that most cliques have only one \node\ in common, if
any. This is responsible for both the inaccuracy of the models
concerning some clusterings and for the irregularities one can observe
on some distributions. Figure~\ref{fig_bip_clustering} plots the
distribution of the overlap between cliques. This overlap is very
small for all random bipartite \networks\ while it is non trivial for
the original \networks.

More precisely, in the case of \Internet, we noticed the presence of a
sub-\network\ of only $94$ \nodes\ which contains all the $494$ cliques of
size $14$ and more. This very dense sub-\network\ implies
that the clustering of each of the $94$ nodes is very high. However,
they have almost no impact on the clustering of the whole \network\
(due to the average). On the other hand, in the $\bot$-projection of
random bipartite networks, these large cliques are disseminated all
over the \network\ which brings two artifacts: there are a lot of
\nodes\ having a degree between $14$ and $29$ which explains the bump
on degree distribution (a similar phenomenon can be observed on \Web),
and the number of \nodes\ with high clustering is drastically
increased because of their presence in large cliques (from $94$ to
$50,000$).

\medskip

These experimental results should also be compared to the ones
obtained with the currently most used models, presented in
Section~\ref{sec_context}.  This comparison gives evidence for the
fact that the models we propose may be considered as an important step
towards the realistic modeling of \cns.

All these remarks hold both for the growing bipartite model and for
the random one. This is worth to notice, since it may be very
important in some contexts that the model produces {\em growing}
graphs with realistic properties, and in other contexts that the
obtained graphs are representative of a precise class of graphs.

\section{\label{sec_conclusion}Conclusion and discussion}

\noindent
In this paper, we propose bipartite graphs as a general tool for
the modeling of real-world \cns. They make it possible to achieve the
following challenges:
\begin{itemize}
\item the obtained networks have the three main wanted properties
(logarithmic average distance,  high clustering and power law degree
distribution),
\item the models are based on a {\em realistic} construction process
representative of what happens in some real-world cases, and
\item their definitions are simple enough to make it possible to give
some  intuition and some proofs of their properties.
\end{itemize}

\noindent
Moreover, they can be derived in two versions: one which relies on
random sampling among a class of graphs, and one which relies on an
iterative construction process. This makes them suitable for a wide
variety of usages.
Moreover, it is very simple to obtain \networks\ using this
model (we provide a generator at \cite{progurl}), which makes it
highly suitable for simulation purposes.

The model is based on the discovery that all real-world \cns\ have an
underlying bipartite structure. Some networks naturally have this
structure and, for the others, we show that they can be decomposed
into cliques which make such a structure emerge. This shows that the
main properties of \cns\ can be viewed as consequences of this
bipartite structure, and that the model captures a general behavior of
complex systems.

\medskip

However, as already stressed in previous sections, the overlapping
between cliques is not taken into account by the bipartite model which
in some way distributes cliques all over the networks independently of
the nodes implied. On the contrary, it seems obvious that \networks\
such as \Actors\ are not randomly constructed: actors from a same
country are more likely to play together, for instance. This lack of overlapping can
also be described on the bipartite \networks: if two top nodes have
more than one bottom node in the intersection of their neighborhood,
then this yields a non trivial bipartite
clique. On the other hand, for the \networks\ generated with both
bipartite models, most of these bipartite cliques are trivial ones (as
long as there are no too many cliques).

An analogy can be made with the clustering in random \networks\ (ER
\networks\ for instance), in which neighborhoods of \nodes\ are very
sparse while real-world neighborhoods are quite dense: one could say that
real-world bipartite networks are bi-clusterized while random ones are
not, even if they capture the most common properties.

\medskip

There are many directions in which this work may be extended. Solving
the previous drawback is one of them. This model might also be
extended to the case of directed and weighted \networks. These
problems rely on giving a new definition to the concept of clique
which can be used in this context.

Another similar problem occurs when the \network\ is only partially
known. In this case, some \links\ are missing, which might yield to
only trivial cliques. A solution to this problem could be to study a
model with quasi-cliques. Embedding this concept in the bipartite
vision however is nontrivial and remains to be done.

One may also use this model to deepen the study of some phenomena of
high interest like the robustness of networks, the spread of rumors
and diseases, etc. The random graph model with prescribed degree
distribution already led to important advances on these questions
\cite{cohen00resilience,cohen01breakdown,New02,PV01}. They should now
be extended to the bipartite models in order to evaluate the impact of
clustering on these problems. We argue that this is a strength of our
approach since results on random graphs with prescribed degrees
can be directly adapted to our model in order to take the clustering
into account.

\medskip

Finally, let us emphasize on the fact that the study of
real-world \cns\ is only at its beginning. The discovery of their
statistical properties, the analysis of the impact of these
properties, their integration into accurate models, and the use of
these models in simulation and analysis are key issues for our
understanding of real-world \cns, which has crucial
fundamental and applicative implications. Our work lies in this
context. It proposes a solution to the problem of the realistic random
modeling of real-world \cns\ (in the sense of the three main observed
properties), and it points out some relevant directions for further
research.

\medskip

\noindent
{\em Acknowledgments.}
We thank Annick Lesne, Cl\'emence Magnien and James Martin
for careful reading of preliminary versions
and useful comments. We also thank the anonymous referees
for helpful comments.

\small
\bibliographystyle{plain}
\bibliography{xbib}

\end{document}